\documentclass[10pt]{article}
\usepackage{epsfig}

\def\to{\rightarrow}

\def\bi{\begin{itemize}}
\def\ei{\end{itemize}}
\def\te{\tilde e}

\def\tb{\tilde b}

\def\ttau{\tilde \tau}

\def\tell{\tilde\ell}

\def\tw{\widetilde W}
\def\tz{\widetilde Z}

\newcommand{\hepph}[1]{hep-ph/#1}
\newcommand{\hepex}[1]{hep-ex/#1}
\newcommand\app[3]{{\it Astropart.\ Phys.\ }{\bf #1} (#2) #3}
\newcommand\ijmpa[3]{{\it Int.\ J.\ Mod.\ Phys.\ }{\bf A #1} (#2) #3}
\newcommand\jhep[3]{{\it J. High Energy Phys.\ }{\bf #1} (#2) #3}
\newcommand\prl[3]{{\it Phys.\ Rev.\ Lett.\ }{\bf #1} (#2) #3}
\newcommand\plb[3]{{\it Phys.\ Lett.\ }{\bf B #1} (#2) #3}
\newcommand\npb[3]{{\it Nucl.\ Phys.\ }{\bf B #1} (#2) #3}
\newcommand\prd[3]{{\it Phys.\ Rev.\ }{\bf D #1} (#2) #3}
\newcommand\prep[3]{{\it Phys.\ Rept.\ }{\bf #1} (#2) #3}
\newcommand\pr[3]{{\it Phys.\ Rev.\ }{\bf #1} (#2) #3}

\begin{document}

\title{Updated constraints on the minimal supergravity model}

\author{
H. Baer, C. Bal\'azs\footnote{Talk given by C. Bal\'azs at SUSY02.}, A. Belyaev
{\it (Department of Physics, Florida State University)} \\
J.~K. Mizukoshi
{\it (Instituto de F\'{\i}sica, Universidade de S\~ao Paulo)} \\
X. Tata, Y. Wang 
{\it (Department of Physics and Astronomy, University of Hawaii)} \\
}
\date{}
\maketitle

\begin{abstract}
Recently, refinements have been made on both the
theoretical and experimental determinations of 
{\it i}.) the mass of the lightest Higgs scalar,
{\it ii}.) the relic density of cold dark matter in the universe,
{\it iii}.) the branching fraction for the radiative $b \to s \gamma$ decay,
{\it iv}.)  the muon anomalous magnetic moment, and
{\it v}.) the flavor violating decay $B_s\to \mu^+\mu^-$. 
In this work, we present constraints from each of these quantities
on the minimal supergravity model as embedded in the updated version of the
computer program ISAJET v7.64. 
Improvements and updates since our published work are especially emphasized.
The combination of constraints
points to certain favored regions of model parameter space where
collider and non-accelerator SUSY searches may be more focused.
\end{abstract}

\section{Introduction}
\label{sec:intro}

Particle physics models including
supersymmetry solve a host of problems occurring in non-supersymmetric
theories, and predict a variety of new matter states--- the sparticles---
at or around the TeV scale%
\cite{reviews}. 
%
The so-called {\it minimal} supergravity (mSUGRA) model (sometimes also
referred to as the CMSSM) has traditionally been the most popular
choice for phenomenological SUSY analyses. In mSUGRA, it is assumed that
the Minimal Supersymmetric Standard Model (MSSM) is valid from the weak
scale all the way up to the GUT scale $M_{GUT}\simeq 2\times 10^{16}$
GeV, where the gauge couplings $g_1$ and $g_2$ unify. In many of the
early SUGRA models\cite{sugra}, a simple choice
of K\"ahler metric $G_i^j$ and gauge kinetic function $f_{AB}$ led to
{\it universal} soft SUSY breaking scalar masses ($m_0$), gaugino masses
($m_{1/2}$) and $A$-terms ($A_0$) at $M_{GUT}$.  This assumption of
universality in the scalar sector
leads to the phenomenologically required suppression of
flavor violating processes that are supersymmetric in origin.  
%
In the mSUGRA model, we thus assume universal scalar masses, gaugino
masses (as a consequence of assuming grand unification) and $A$-terms. 
We will also require that electroweak symmetry is
broken radiatively (REWSB), allowing us to fix the magnitude,
but not the sign, of the superpotential Higgs mass term $\mu$ so as to
obtain the correct value of $M_Z$. Finally, we
trade the bilinear soft
supersymmetry breaking (SSB) parameter $B$ for $\tan\beta$ (the ratio of
Higgs field vacuum expectation values).  Thus,
the parameter set
\begin{equation}
m_0,\ m_{1/2},\ A_0,\ \tan\beta ,\ \ {\rm and}\ \ sign(\mu )
\end{equation}
completely determines the spectrum of supersymmetric matter and physical 
Higgs fields.

In our calculations, we use ISAJET v7.64 \cite{isajet} since this version 
includes a number of improvements in calculating the SUSY particle mass spectrum
compared to v7.58 used in Ref.\cite{Baer:2002gm}.
Once the SUSY and Higgs masses and mixings are known, then a host of
observables may be calculated, and compared against experimental
measurements. The most important of these include: 
\begin{itemize}
\vspace{-2.6mm}
\item lower limits on sparticle and Higgs boson masses from new particle
searches at LEP2,
\vspace{-2.6mm}
\item the relic density of neutralinos originating from the Big Bang,
\vspace{-2.6mm}
\item the branching fraction of the flavor changing decay $b\to
s\gamma$,
\vspace{-2.6mm}
\item the value of muon anomalous magnetic moment $a_\mu = (g-2)_\mu/2$ and
\vspace{-2.6mm}
\item the lower bound on the rate for the rare decay $B_s\to\mu^+\mu^-$.
\end{itemize}
\vspace{-2.6mm}
Our goal is to delineate the mSUGRA parameter space region consistent
with all these constraints.  
%
In our analysis,
we incorporate a new calculation of the neutralino relic density
that has recently become available\cite{bbb}.  
We also present improved $b\to s\gamma$ branching fraction predictions in
accord with the current ISAJET release. We discuss constraints
imposed by the measurement of the muon anomalous magnetic moment.
Finally, we delineate
the region of mSUGRA parameter space excluded by the CDF lower
limit\cite{cdf} on the branching fraction of $B_s\to \mu^+\mu^-$.  This
constraint is important 
for very large 
$\tan\beta$'s\cite{bmm}.

Within the mSUGRA framework, the parameters $m_0$ and $m_{1/2}$ are the
most important for fixing the scale of sparticle masses. The
$m_0$-$m_{1/2}$ plane (for fixed values of other parameters) is convenient
for
a simultaneous display of these constraints, and hence, of parameter
regions in accord with all experimental data.
  

\section{Constraints and calculations in the mSUGRA model}
\label{sec:constraints}

{\bf ~~~ Constraints from LEP2 searches}

Based on negative searches for superpartners at LEP2,
we require 
\begin{itemize}
\vspace{-2.6mm}
\item $m_{\tw_1}>103.5$ GeV ~~~ and ~~~ $m_{\te_{L,R}}>99$ GeV 
provided $m_{\tell}-m_{\tz_1}>10$ GeV, 
\end{itemize}
\vspace{-2.6mm}
which is the most stringent of the slepton mass limits.
The LEP2 experiments also set a limit on the SM Higgs boson mass:
$m_{H_{SM}}>114.1$ GeV\cite{lep2_h}. In our mSUGRA parameter space scans, the
lightest SUSY Higgs boson $h$ is almost always SM-like. The exception occurs
when the value of $m_A$ becomes low at very large values 
of $\tan\beta$. For clarity, we show contours where
\bi
\vspace{-2.6mm}
\item $m_h>114.1$ GeV,
\vspace{-2.6mm}
\ei
and will direct the reader's attention to any regions where this bound might
fail. \\

{\bf Neutralino relic density}

Measurements of galactic rotation curves, binding of galactic clusters, 
and the large scale structure of the universe all point to the need for
significant amounts of cold dark matter (CDM) in the universe. In addition, 
recent measurements of the power structure of the cosmic
microwave background, and measurements of distant supernovae, point
to a cold dark matter density\cite{cdm}
\bi
\vspace{-2.6mm}
\item $0.1 <\Omega_{CDM}h^2<0.3$.
\vspace{-2.6mm}
\ei
The lightest neutralino of mSUGRA is an excellent candidate for
relic CDM particles in the universe.
The upper limit above represents a true constraint, while the 
corresponding lower
limit is flexible, since there may be additional sources of CDM such
as axions, or states associated with the hidden sector and/or extra
dimensions.

To estimate the relic density of neutralinos in the mSUGRA model,
we use the recent calculation in Ref. \cite{bbb}. 
In that work, 
all relevant neutralino annihilation and co-annihilation
reactions are evaluated at tree level using the CompHEP\cite{comphep}
program.
The annihilation cross section times velocity is relativistically thermally
averaged\cite{graciela}, 
which is important for obtaining the correct neutralino relic 
density in the vicinity of annihilations through $s$-channel resonances. \\

{\bf The $b\to s\gamma$ branching fraction}

The branching fraction $BF(b\to s\gamma )$ has recently been measured by
the BELLE\cite{belle}, CLEO\cite{cleo} and ALEPH\cite{aleph}
collaborations.  Combining statistical and systematic errors in
quadrature, these measurements give $(3.36\pm 0.67)\times 10^{-4}$
(BELLE), $(3.21\pm 0.51)\times 10^{-4}$ (CLEO) and $(3.11\pm 1.07)\times
10^{-4}$ (ALEPH). A weighted averaging of these results yields $BF(b\to
s\gamma )=(3.25\pm 0.37) \times 10^{-4}$. The 95\% CL range corresponds
to $\pm 2\sigma$ away from the mean. To this we should add uncertainty
in the theoretical evaluation, which within the SM dominantly comes from
the scale uncertainty, and is about 10\%.
Together, these imply the bounds,
\bi
\vspace{-2.6mm}
\item $2.16\times 10^{-4}< BF(b\to s\gamma )< 4.34 \times 10^{-4}$.
\vspace{-2.6mm}
\ei
In our study, we show contours of $BF(b\to s\gamma )$ of
2, 3, 4 and $5\times 10^{-4}$.

The calculation of $BF(b\to s\gamma )$ used here is based upon the
program of Ref. \cite{bsg}. 
In our calculations, we also implement the running $b$-quark mass
including SUSY threshold corrections as calculated in ISAJET;
these effects can be important at large values of the 
parameter $\tan\beta$\cite{degrassi}.
Our value of the SM $b\to s\gamma$
branching fraction yields $3.4\times 10^{-4}$, with a scale uncertainty
of 10\%. \\

{\bf Muon anomalous magnetic moment}

The muon anomalous magnetic moment $a_\mu =(g-2)_\mu/2$ has been
recently measured to high precision by the E821 experiment\cite{Bennett:2002jb}:
$ a_\mu=11659204(7)(5)\times 10^{-10}$.
%
%
%
%
The most challenging parts of the SM calculation are the hadronic 
light-by-light\cite{lbl} and vacuum polarization (HVP)\cite{HVPee} contributions 
and their uncertainties. Presently these results are in dispute. In the case of 
the HVP the use of tau decay data can reduce the error, but the interpretation 
of these data is somewhat controversial\cite{HVPtau}.
Thus, the deviation of the measurement from the SM depends on which prediction 
is taken into account. According to the recent analysis by Hagiwara et 
al.\cite{HVPee}:
\bi
\vspace{-2.6mm}
\item $11.5<\delta a_\mu\times 10^{10}<60.7$. 
\vspace{-2.6mm}
\ei
A different assessment of the theoretical uncertainties\cite{HVPee}
using the procedure described in ref.\cite{Baer:2002gm} gives,
\bi
\vspace{-2.6mm}
\item $-16.7< \delta a_\mu\times 10^{10}<49.1$.
\vspace{-2.6mm}
\ei
In view of the
theoretical uncertainty, 
we only
present contours of $\delta a_\mu$, as calculated using the program
developed in \cite{bbft}, and leave it to the reader to decide the
extent of the parameter region allowed by the data. \\

{\bf $B_s\to\mu^+\mu^-$ decay}


The branching fraction of $B_s$ to a pair of muons has been
experimentally
bounded by CDF\cite{cdf}:
\bi
\vspace{-2.6mm}
\item $BF(B_s\to\mu^+\mu^- )< 2.6\times 10^{-6}$.
\vspace{-2.6mm}
\ei
%
A potentially important contribution to this decay is mediated by
the neutral states in the Higgs sector of supersymmetric
models. While this branching fraction is very small within the SM
($BF_{SM}(B_s \to \mu^+\mu^-)\simeq 3.4 \times 10^{-9}$), the amplitude
for the Higgs-mediated decay of $B_s$ grows as $\tan^3\beta$ within the
SUSY framework, and hence can completely dominate the SM contribution if
$\tan\beta$ is large.  
In our analysis we use the results from the last paper in Ref.\cite{bmm} 
to delineate 
the region of mSUGRA parameters excluded by the CDF upper limit on its 
branching fraction. 


\section{Results}
\label{sec:results}

To generate numerical results, in this work we use ISAJET v7.64 that includes 
several improvements over v7.58 which was used in Ref.\cite{Baer:2002gm}. These 
changes lead to important differences in the figures when compared
with Ref.\cite{Baer:2002gm}. Notably, 
the boundary of the region excluded by the lack of REWSB moved to higher 
$m_{0}$ values and the allowed relic density region along this boundary changed,
especially for the lower $\tan\beta$ values.
Furthermore, 
for high $\tan\beta$ and $\mu < 0$ the diagonal corridors allowed by the 
relic density are considerably shifted and narrowed.
Finally, the area allowed by relic density near the boundary of the stau LSP 
region shrank at low $\tan\beta$'s.

\begin{figure}
\epsfig{file=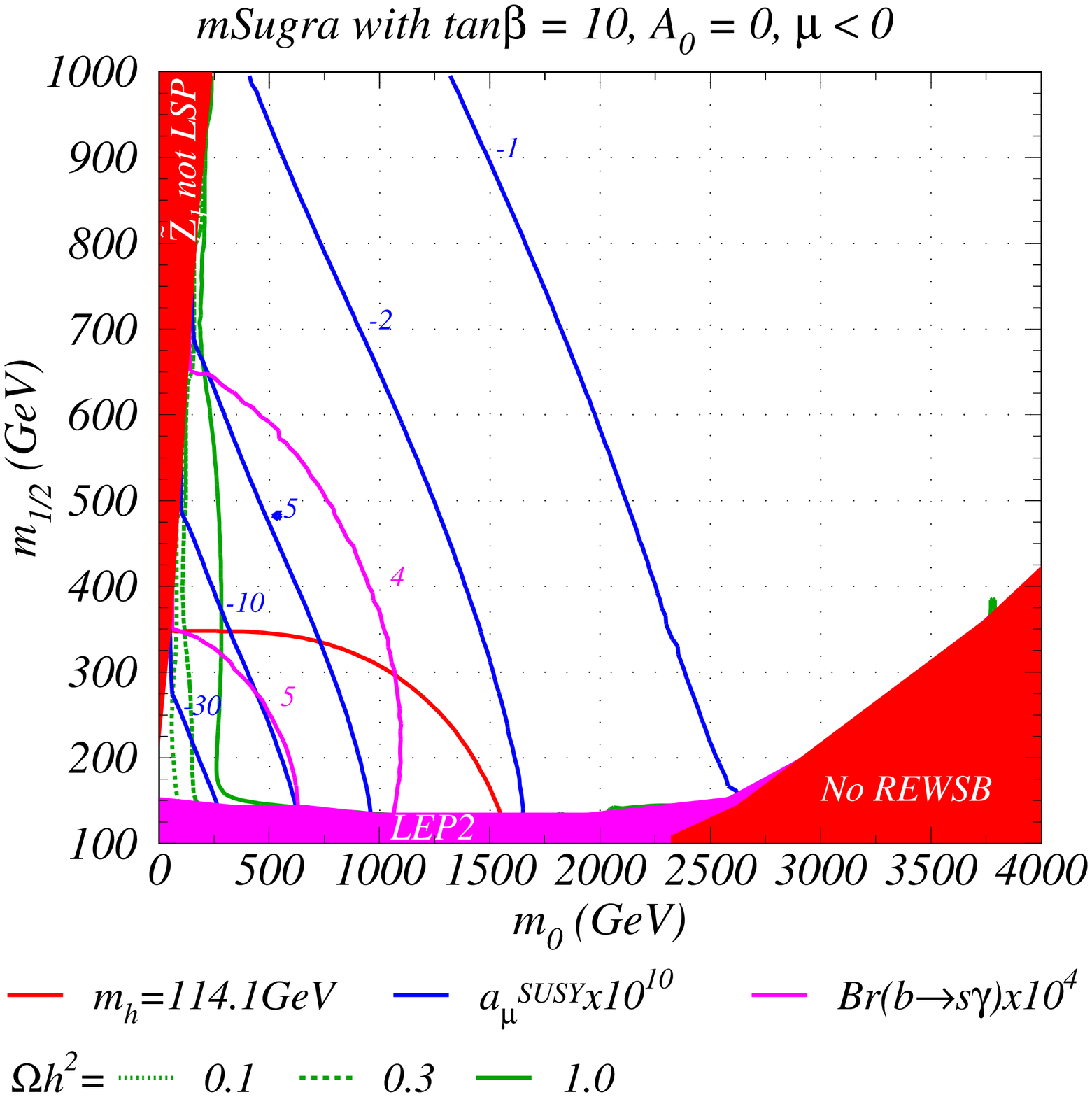,width=8cm} 
\epsfig{file=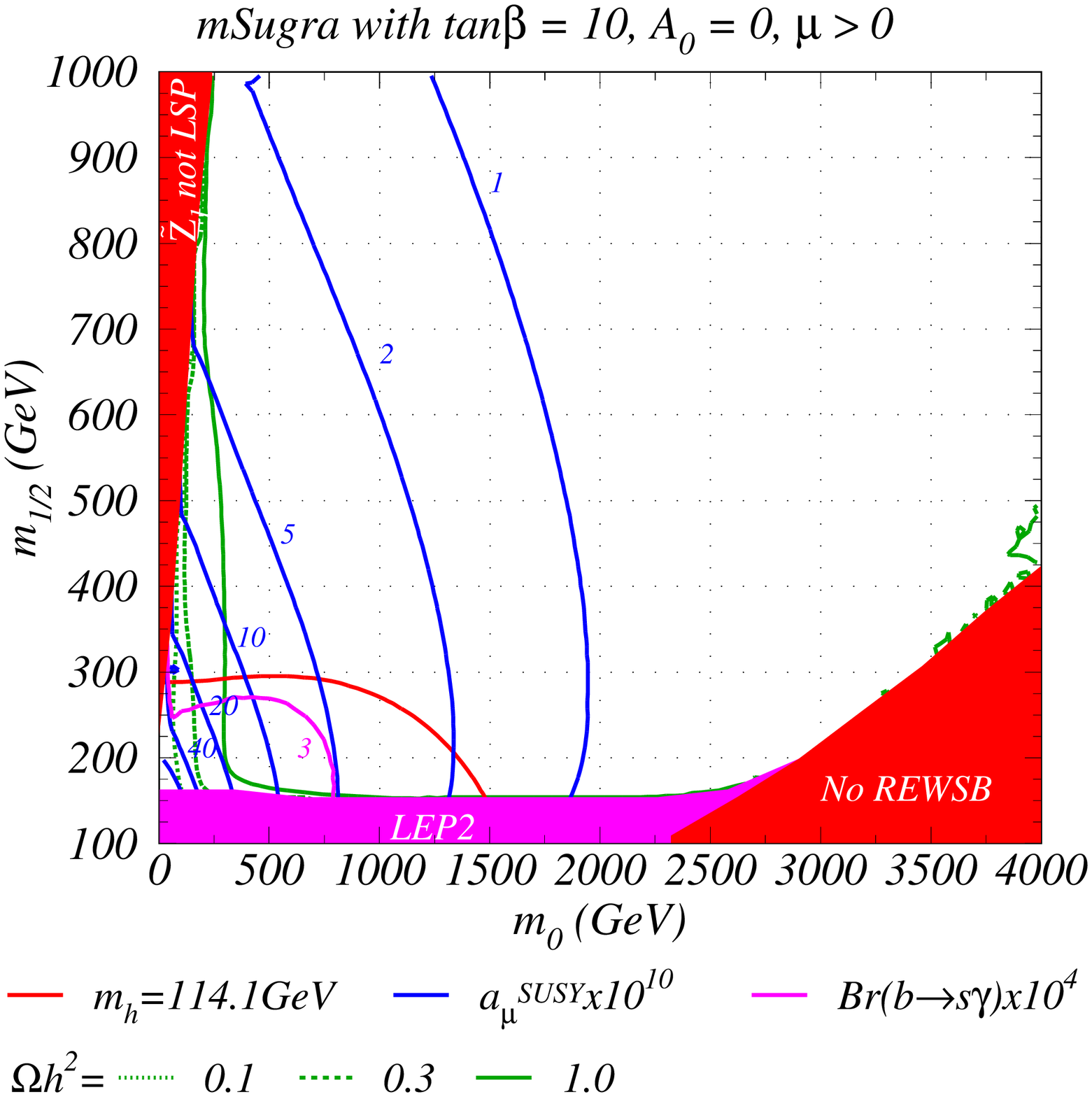,width=8cm} 
\caption{Plot of constraints for the mSUGRA model in the 
$m_0\ vs.\ m_{1/2}$ plane for $\tan\beta =10$ and $A_0=0$.
We plot contours of the CDM relic density, $m_h=114.1$ GeV, 
the muon anomalous magnetic moment $a_\mu$ ($\times 10^{10}$) and
contours of $b\to s\gamma$ branching fraction ($\times 10^{4}$).}
\label{fig:sug10}
\end{figure}

Our first results are plotted in Fig.\ref{fig:sug10}. Here, we
show the $m_0\ vs.\ m_{1/2}$ plane for $A_0=0$, $\tan\beta =10$ and
both signs of $\mu$.
The red shaded regions are excluded either due to a
lack of REWSB (right-hand side), or a stau LSP (left-hand side).
The magenta region is excluded by searches  for charginos and
sleptons at LEP2. The region below the red contour is excluded by LEP2
Higgs searches, since here $m_h<114.1$ GeV. In addition, we show
regions of neutralino relic density with green contours marking
$\Omega_{\tz_1} h^2 =0.1$ (dotted), 0.3 (dashed) and 1.0 (solid). 
The region right to the solid green contour has 
$\Omega_{\tz_1} h^2>1$, 
and would thus be excluded since the age of the universe would be less
than 10 billion years. 
There is no constraint arising from $B_s\to \mu^+\mu^-$ decay at 
$\tan\beta =10$.

For $\mu <0$ the magenta contours denote values of
$BF(b\to s\gamma )= 4$ and $5\times 10^{-4}$ and the blue
contours denote values of $\delta a_\mu =-30, -10, -5, -2$ and 
$-1\times 10^{-10}$, moving from lower left to upper right.
An intriguing feature of the plot is that the region with the allowed 
relic density in the lower left part, where neutralinos mainly annihilate
via $t$-channel slepton exchange 
to lepton-anti-lepton pairs
is essentially excluded by the $m_h$, $b\to s\gamma$ and
$\delta a_\mu$ constraints. That leaves two
allowed regions
with a preferred relic density: one that runs near the stau LSP region, where
$\ttau_1-\tz_1$ co-annihilation effects reduce an otherwise
large relic density 
(as pointed out by Ellis {\it et al.}\cite{ellis_co}).
This region has a highly fine-tuned relic density, since a slight change in
$m_0$ leads to either too light or too heavy of a $\ttau_1$ mass
to give $0.1<\Omega h^2<0.3$\cite{ellis_ft,bbb}. 
The other runs parallel
to the REWSB excluded region for $m_{1/2} > 400$ GeV in the ``focus point'' SUSY 
region. It occurs when the $\tz_1$ has a sufficiently
large higgsino component that annihilation into $WW$, $ZZ$ and $Zh$
pairs reduces the relic density\cite{feng_relic,bbb}. 

For $\mu >0$ 
almost the entire plane shown is in
accord with the measured branching fraction of $b\to s\gamma$. 
The blue contours denote values of $\delta a_\mu =60$, 40, 20, 10, 5, 2 and
$1\times 10^{-10}$. Constraints from $\delta a_{\mu}$ as
well as from $B_s \to \mu^+\mu^-$ are not relevant for this case.
%
In this case 
the slepton annihilation region
of relic density has a small surviving region just beyond 
the Higgs mass contour. 
For the most part, to attain a 
preferred value of neutralino relic density, one must again live in the 
stau co-annihilation
region. A final 
possibility is to be in the slepton annihilation region, but then the 
value of $m_h$ should be slightly beyond the LEP2 limit; 
in this case,
a Higgs boson signal may be detected in Run 2 of the Fermilab
Tevatron\cite{run2}.

\begin{figure}
\epsfig{file=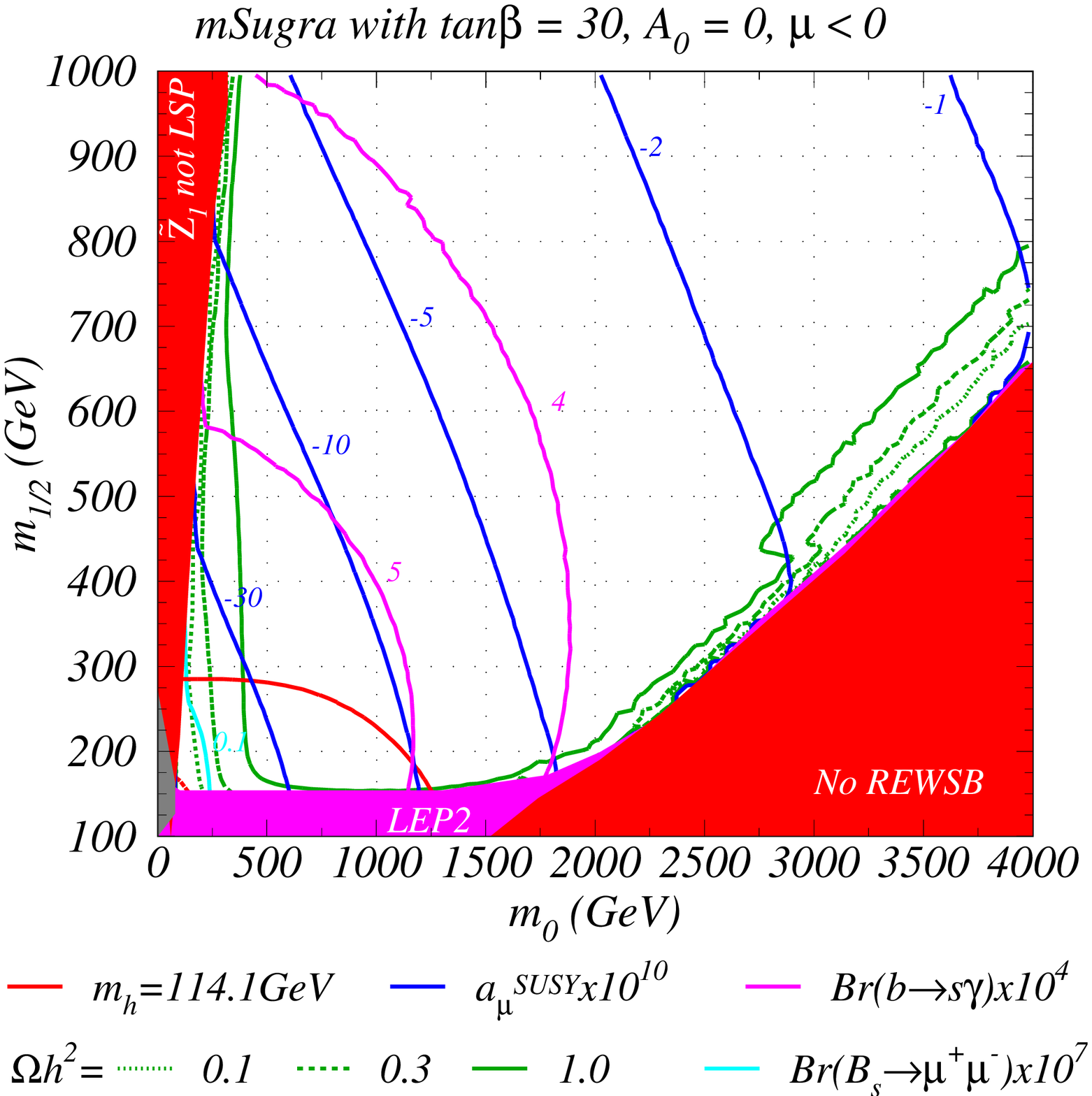,width=8cm} 
\epsfig{file=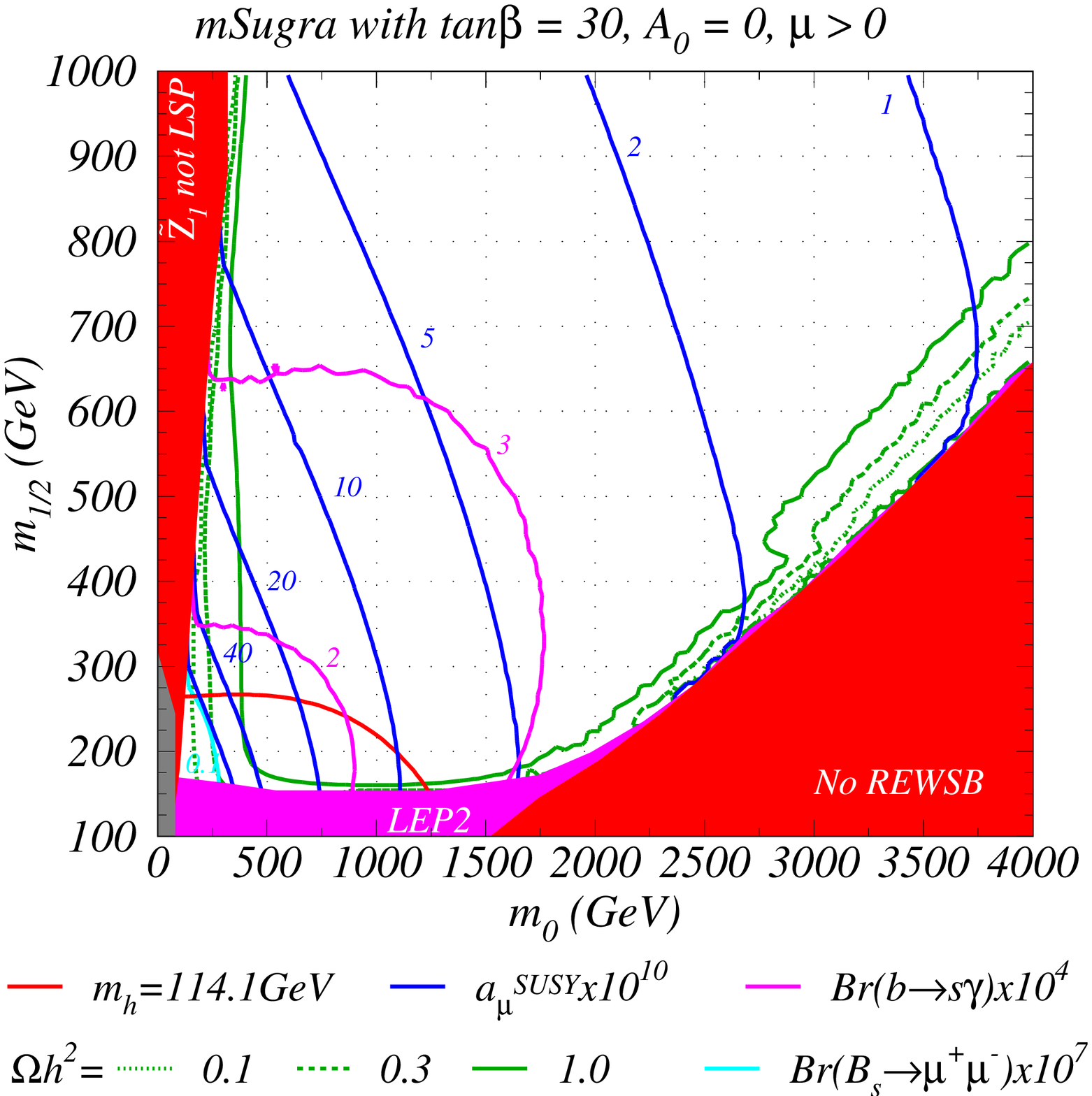,width=8cm} 
\caption{Same as Fig. \ref{fig:sug10}, but for $\tan\beta =30$. 
The light blue contour labeled 0.1 denotes where $B(B_s \to
\mu^+\mu^-)= 0.1 \times 10^{-7}$. In subsequent figures these branching
fractions contours are all labeled in units of $10^{-7}$. }
\label{fig:sug30}
\end{figure}

We next turn to our results for $\tan\beta=30$ shown in Fig.\ref{fig:sug30}.
The gray region in the bottom left corner of the plot is
excluded because $m_{\ttau_1}^2 < 0$. In this case, the allowed
region of the relic density in the lower-left has expanded considerably
owing to enhanced neutralino annihilation to $b\bar{b}$ and
$\tau\bar{\tau}$ at large $\tan\beta$.  Both lighter values of
$m_{\ttau_1}$ and $m_{\tb_1}$ and also large $\tau$ and $b$ Yukawa
couplings at large $\tan\beta$ enhance these $t$-channel annihilation
rates through virtual staus and sbottoms. Unfortunately, for $\mu < 0$ 
the region excluded by $BF(b\to s\gamma )$ and by $\delta a_\mu$ 
also expands, and most of the cosmologically preferred region is
again ruled out.
As before, we are left with the corridors of stau
co-annihilation and an enlarged focus point scenario\cite{feng_relic,bbb} 
as the only surviving regions.
 
For 
$\mu >0$
the magenta contours of
$BF(b\to s\gamma )$ correspond to $2$ and $3\times 10^{-4}$. Thus, the lower
left region is excluded since it leads to too {\it low} a value
of $BF(b\to s\gamma )$. The $\delta a_\mu $ contours begin from lower left 
with $60\times 10^{-10}$, then proceed to 40, 20, 10, 5, 2 and $1\times 10^{-10}$.
A fraction of the slepton annihilation region of relic density is
excluded also by too large a value of $\delta a_\mu$. Of course,
a reasonable relic density may also be achieved in the stau co-annihilation 
and focus point regions of parameter space.

\begin{figure}
\epsfig{file=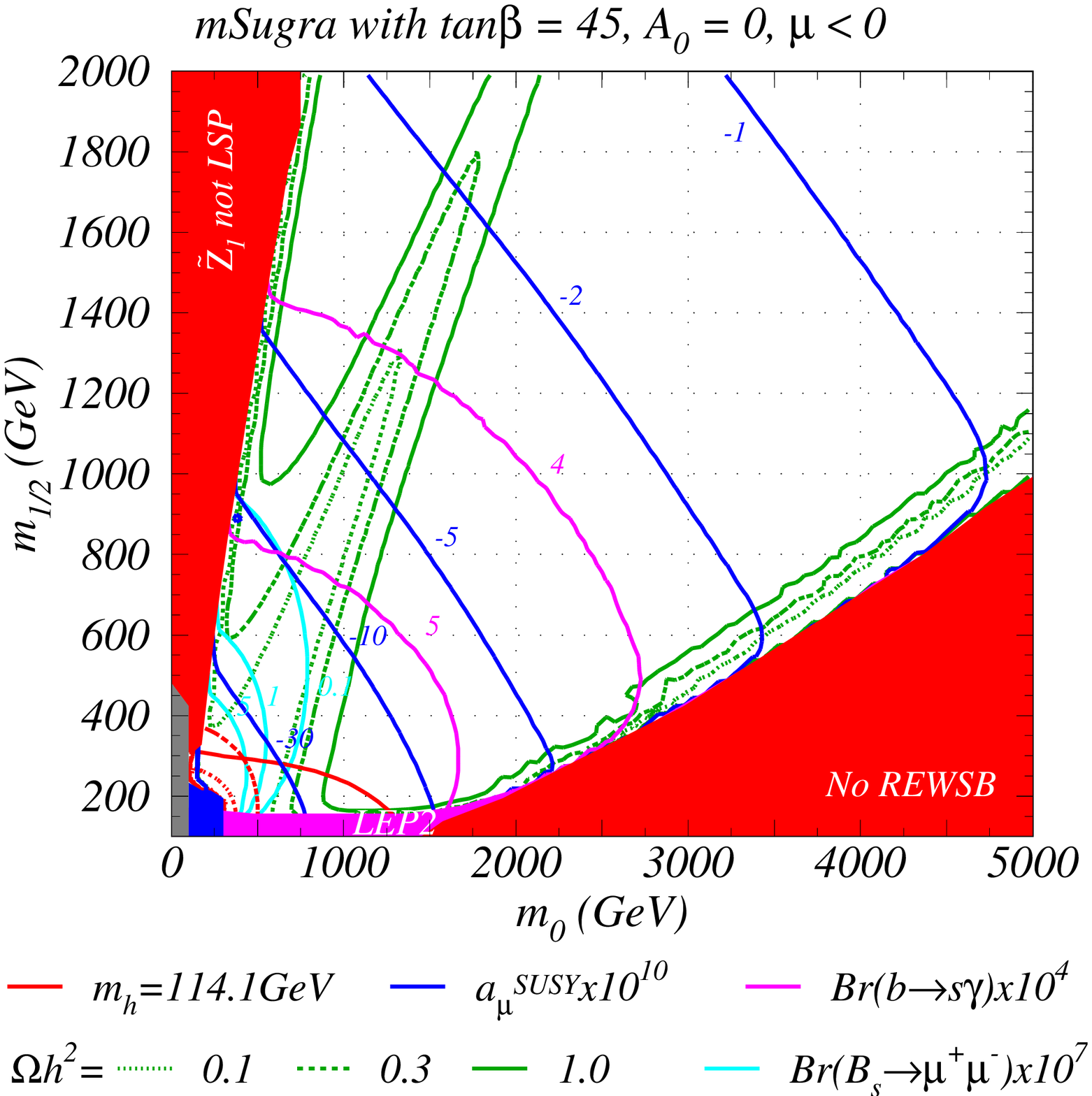,width=8cm} 
\epsfig{file=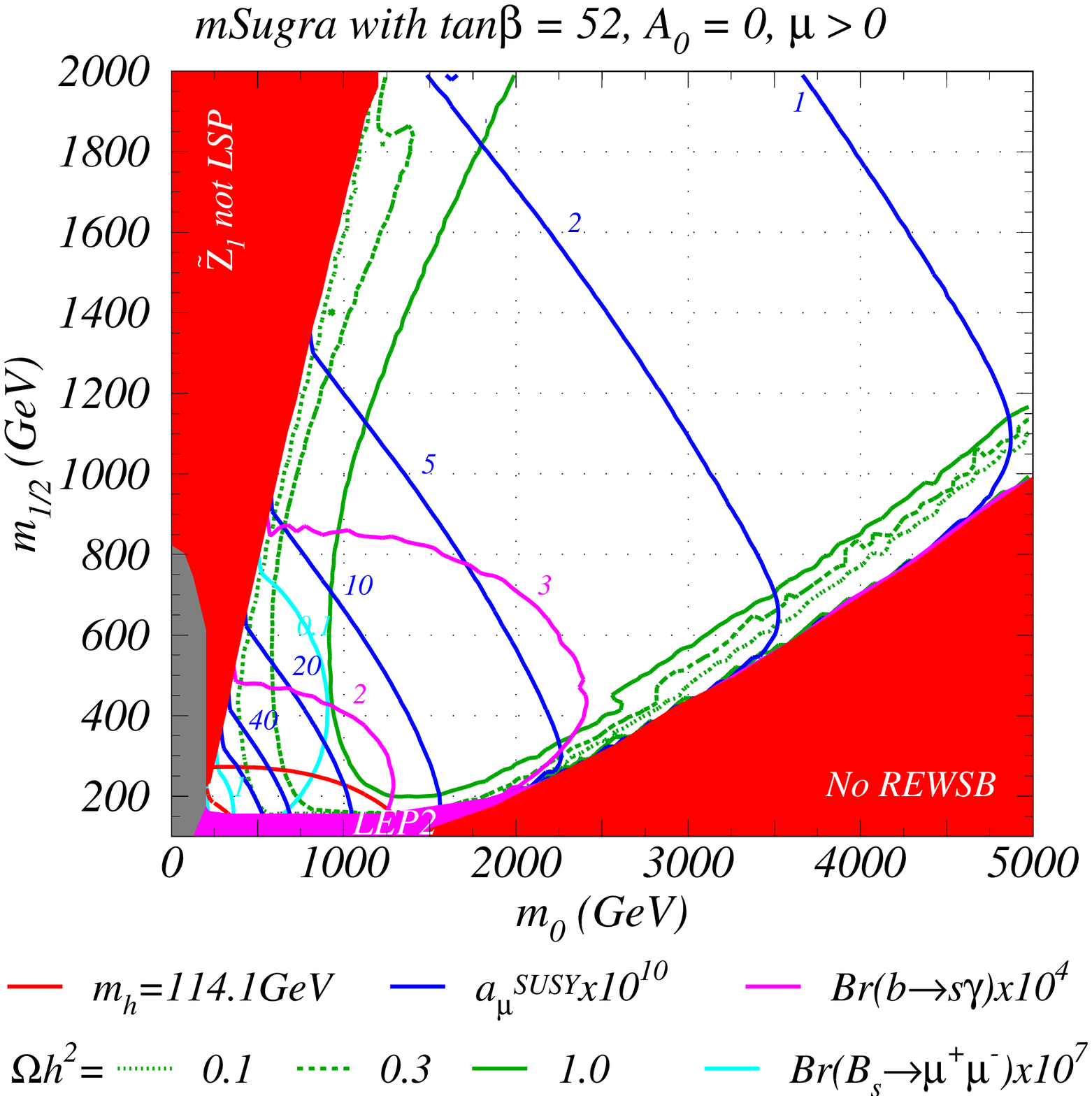,width=8cm} 
\caption{Same as Fig. \ref{fig:sug10}, but for $\tan\beta =45$, $\mu <0$
and for $\tan\beta =52$, $\mu >0$.
The inner and outer red dashed lines are contours of $m_A=100$ and 
$m_A= 200$~GeV, respectively.}
\label{fig:sug4552}
\end{figure}

Next, we turn to Fig.\ref{fig:sug4552} where we examine 
the mSUGRA parameter plane for very large values of
$\tan\beta =45$ and $\mu <0$. 
The gray and red regions are as in previous figures. The
blue region is excluded because $m_A^2<0$, denoting again a lack of
appropriate REWSB. The inner and outer red dashed lines are contours of
$m_A=100$ and $m_A= 200$~GeV, respectively. The former is roughly the
lower bound on $m_A$ from LEP experiments. In between these contours,
$h$ is not quite SM-like, and the mass bound from LEP may be somewhat lower
than $m_h=114.1$~GeV shown by the solid red contour, but outside the
200~GeV contour this bound should be valid.
%
Much of the lower-left region is excluded by too high a
value of $BF(b\to s \gamma )$ and too low a value of $\delta a_\mu$.  In
addition, in this plane, the experimental limit on $B_s\to\mu^+\mu^-$
enters the lower-left, where values exceeding $26\times 10^{-7}$ are
obtained.
It seems that in the upper region which is  favored by the $b \to s\gamma$
constraint, detection of $B_s \to \mu^+\mu^-$ at the Tevatron will be
quite challenging.

In this figure, the relic density regions are qualitatively
different
from the lower $\tan\beta$ plots. A long diagonal strip running from
lower-left to upper-right occurs because in this region, neutralinos
annihilate very efficiently through $s$-channel $A$ and $H$ Higgs
graphs, where the total Higgs widths are very large due to the large $b$
and $\tau$ Yukawa couplings for the high value of $\tan\beta$ in this
plot. Adjacent to this region allowed regions where
neutralino annihilation is still dominated by the $s$-channel Higgs
graphs, but in this case the annihilation is somewhat off-resonance. The
$A$ and $H$ widths are so large
that even if
$|2m_{\tz_1}-m_{A(H)}|$ is relatively large, efficient annihilation can
still take place. (An improvement of the Higgs widths is adopted for these
plots compared to Ref.\cite{Baer:2002gm}.)

For the case of $\mu >0$, 
we show 
the mSUGRA parameter space plane for $\tan\beta =52$. 
In this plane, the relic density annihilation corridor 
occurs near the boundary of the excluded $\ttau_1$ LSP region. 
The width of the $A$ and $H$ Higgs scalars is very wide, 
so efficient $s$-channel annihilation through the Higgs poles can occur
throughout much of the allowed parameter space. But the 
annihilation is not overly efficient due to the large breadth of the 
Higgs resonances.
%
In much of the region with $m_{1/2}<400$ GeV, the value of
$BF(b\to s\gamma )$ is below
$2\times 10^{-4}$, so that some of the lower allowed relic density region
where annihilation occurs through $t$-channel stau exchange is excluded.
In contrast, the value of $\delta a_\mu$ is 
in the range of $10-40\times 10^{-10}$, which is in accord with
the E821 measurement. The value of $m_h$ is almost always
above 114.1 GeV, and the $BF(B_s\to\mu^+\mu^- )$ is always
below $10^{-7}$, and could (if at all) be detected with several years of
main injector operation. 
%


In conclusion, we have presented updated constraints on the mSUGRA model
from {\it i.}) the LEP2 constraints on sparticle and Higgs boson masses, 
{\it ii.}) the neutralino relic density $\Omega_{\tz_1}h^2$, 
{\it iii.}) the branching fraction $BF(b\to s\gamma )$, 
{\it iv.}) the muon anomalous magnetic moment $a_\mu$ and
{\it v.}) the leptonic decay $B_s\to\mu^+\mu^-$. Putting all five
constraints together, we find favored regions of parameter space which
may be categorized by the mechanism for annihilating relic neutralinos
in the early universe:
\bi
\vspace{-2.6mm}
\item {\bf 1.} annihilation through $t$-channel slepton exchange
(low $m_0$ and $m_{1/2}$),
\vspace{-2.6mm}
\item {\bf 2.} the stau co-annihilation region 
(very low $m_0$ but large $m_{1/2}$),
\vspace{-2.6mm}
\item {\bf 3.} the focus point region (large $m_0$ but low to 
intermediate $m_{1/2}$) and
\vspace{-2.6mm}
\item {\bf 4.} the flanks of the neutralino $s$-channel annihilation via
$A$ and $H$ corridor at large $\tan\beta$ when $\Gamma_A$ and $\Gamma_H$
are very large.
\vspace{-2.6mm}
\ei

To summarize, we find the five constraints considered in this work
to be highly restrictive. Together, they rule out large regions of parameter
space of the mSUGRA model, including much of the region where
$t$-channel slepton annihilation of neutralinos occurs in the early universe.
The surviving regions {\bf 1.}-{\bf 4.} have distinct characteristics of their
SUSY spectrum, and should lead to distinctive SUSY signatures at colliders.

\section*{Acknowledgments}
 
This research was supported in part by the U.S. Department of Energy
under contracts number DE-FG02-97ER41022 and DE-FG03-94ER40833, and by 
Funda\c{c}\~ao de Amparo \`a Pesquisa do Estado de S\~ao Paulo
(FAPESP).
	


\end{document}